\documentstyle[aps,eqsecnum,preprint]{revtex}
\begin{document}
\draft
\tightenlines

\title{Comprehensive theory of the relative phase 
in  atom-field interactions}
\author{J. Delgado, E. C. Yustas, L. L. S\'anchez-Soto}
\address{Departamento de \'{O}ptica, Facultad de Ciencias
F\'{\i}sicas, Universidad Complutense, 28040  Madrid, Spain}
\author{A. B. Klimov}
\address{Departamento de F\'{\i}sica, Universidad de Guadalajara,  
Revoluci\'on 1500, 44420 Guadalajara, Jalisco, M\'exico}
\date{\today}

\maketitle

\begin{abstract}
We explore the role played by the quantum relative 
phase in a well-known model of atom-field interaction,
namely, the Dicke model.  We introduce an appropriate 
polar decomposition of the atom-field relative amplitudes 
that leads to a truly Hermitian relative-phase operator, 
whose eigenstates correctly describe the phase properties, 
as we demonstrate by studying the positive operator-valued
measure derived from it. We find the probability distribution 
for this relative phase and, by resorting to a numerical 
procedure, we study its time evolution.
\end{abstract}


\narrowtext

\section{Introduction}

The problem of the interaction of  an atomic 
system with a radiation field is a keystone of
quantum optics. Needless to say, it is impossible
to obtain exact solutions to this problem 
and some approximations must be used; 
the most common being that the radiation
field is quasimonochromatic and its frequency
coincides almost exactly with one of the transition
frequencies of the atoms (supposed identical and 
with no direct interaction between them).

The two-level atom is the natural consequence
of this hypothesis~\cite{Allen87}. Such an 
object is  an important tool because it allows us 
to describe  the matter-field interaction in a very 
simple way, and the results constitute a first step 
to deal with more realistic situations that 
could include losses, broadening of the 
atomic lines, etc. To put it bluntly, we can replace 
the whole atomic system by an effective 
two-level system that accounts for all the 
relevant details of  the interaction. The Dicke 
model~\cite{Dicke54,Tavis68}, describing  
the interaction of $A$ identical two-level atoms 
with a single-mode field in a perfect cavity, 
is perhaps the archetype of this situation.

In the semiclassical version of this Dicke model,  
correlations are safely ignored and the field is 
interpreted to be a purely classical electric 
field~\cite{Stroud70,Kumar70,Cives99}. Such an 
approximation has proven to be very successful 
and has the virtue of reducing the problem to the
exclusive knowledge of the atomic dynamics, 
which is studied in terms of the inversion and 
the components of the atomic dipole in-phase
and in-quadrature with the field (i.e., the Bloch
vector). These quadratures are the dispersive 
and absorptive components of the dipole moment 
effective in coupling to the field.

For some phenomena, such as spontaneous
emission by a fully excited atomic system,
the quantization of the field is required. Then, 
one must take care also of the evolution of 
the field  amplitudes, but the atomic dynamics 
is still explained in terms of inversion and dipole 
quadratures.

Classically,  the interaction of matter with light 
is usually treated within the framework of
the Lorentz model~\cite{Meystre91}, which 
assumes that each electron-ion pair behaves 
as a single harmonic oscillator that couples
to the field through its electric dipole
moment. In spite of its simplicity, it is
extraordinarily helpful in developing an
intuitive feeling for the physical mechanisms
involved. Although the dynamics of this model 
is sometimes expressed in quadrature 
components~\cite{Allen87}, the natural 
way of understanding its resonant behavior 
is in terms of  the relative phase between the 
field and the atomic  dipole~\cite{Ashcroft96}.

While the quantum quadratures are well
known, and so are the associated eigenstates,
the operator for this relative phase has 
resisted a quantum description. At this 
respect, we think that, in spite of its 
maturity and success, the Dicke model is 
apparently incomplete since it lacks a 
satisfactory description in terms of this 
relative phase, indispensable to compare 
with the classical world. The main goal of 
this paper is precisely the general description 
of that variable.

When focusing on the relative phase 
between two subsystems, we think the
best way to proceed, much in the spirit of
our previous work in the subject~\cite{shg}, is 
to try  a polar decomposition of the quantum 
amplitudes, which parallels as much as possible 
the  corresponding classical factorization. For the
relative phase between two harmonic 
oscillators, this is a quite straightforward 
procedure and leads to a unitary solution~\cite{relpha}.

For the Dicke model, this polar decomposition
seems to be more involved, mainly because,
unlike for the case of two harmonic oscillators,
the Hamiltonian cannot be cast in terms of 
su(2) operators, but rather in terms of some
polynomial deformation of su(2). These
nonlinear algebras have been examined 
very recently in quite different physical 
contexts~\cite{deformed}, and, by exploiting 
these results, it is  possible to perform such 
a decomposition in an elegant way,  obtaining 
a \textit{bona fide} Hermitian operator 
representing the relative phase we wish to examine.

In this paper we use this operator to introduce 
the associated probability distribution and, then, 
the most relevant dynamical features of the 
Dicke model can be easily explained  using 
this relevant variable.

\section{Quantum dynamics of the Dicke model}

The Dicke model describes the interaction of a
collection of $A$ identical two-level atoms with 
a  quantum single-mode field in a lossless cavity. 
The spatial dimensions of the atomic system are
smaller than the wavelength of the field, so all the
atoms feel the same field. However, the model
neglects the dipole-dipole interaction between atoms
(i.e., their wavefunctions do not overlap in the evolution). 

The Hamiltonian for this model in the electric-dipole 
and rotating-wave approximations reads as 
(in units $\hbar =1$) 
\begin{equation}
H = H_0 + H_{\mathrm{int}}  ,
\end{equation}
with
\begin{eqnarray}
\label{Hamex}
H_0 & = &  \omega_{\mathrm{f}} N , \nonumber \\
& & \\
H_{\mathrm{int}} & = & \Delta \ S_3 + 
g \left ( a^\dagger S_-  + a S_ + \right ) . 
\nonumber 
\end{eqnarray}
Here 
\begin{equation}
N = a^\dagger a + S_3 
\end{equation}
is the excitation number operator,  $g$ is the 
coupling constant (which in this approximation is 
the same for all the atoms and can be chosen as real),
and $\Delta = \omega_{\mathrm{f}} - 
\omega_{\mathrm{a}}$ is the detuning between
the atomic and field frequencies. The field mode 
is described by the annihilation and creation operators 
$a$ and $a^\dagger$, while the collective atomic 
operators are defined by 
\begin{equation}
S_{\pm, 3}  =  
\sum_{j=1}^A \sigma _{\pm, 3}^{j} , 
\end{equation}
and obey the commutation relations
\begin{equation}
\label{ccrsu2}
[S_3, S_\pm ] = \pm S_\pm ,
\qquad 
[S_+, S_- ]  =   2 S_3 .  
\end{equation}

Since all the atoms have the same coupling
constant, we need to consider only symmetrical
atomic states. Then, let us introduce the atomic
Dicke states as~\cite{Dicke54}
\begin{equation}
| M \rangle_{\mathrm{a}} = 
\sqrt{\frac{M! (A-M)!}{A!}}
\sum_p |+ \rangle_{j_1} \ldots
|+ \rangle_{j_M} |- \rangle_{j_{M+1}} 
\ldots | - \rangle_{j_A} , 
\end{equation}
where $| \pm \rangle_j$ are the eigenstates
of the $j$th  atom and the sum runs over
all possible manners of choosing $M$ 
indistinguishable atoms from the group of
$A$ atoms. In the space spanned by these 
Dicke states, the action of the collective 
atomic operators is 
\begin{eqnarray}
S_+ \ | M \rangle_{\mathrm{a}} & = & 
\sqrt{( M+1)(A-M)}\  | M +1 \rangle_{\mathrm{a}} ,  
\nonumber \\
S_- \ | M \rangle_{\mathrm{a}} & = &  
\sqrt{M (A - M +1)} \ | M -1 \rangle_{\mathrm{a}} , \\
S_3 \ | M \rangle_{\mathrm{a}} & = & 
( M - A/2 ) \ | M \rangle_{\mathrm{a}}   , \nonumber 
\end{eqnarray}
where the label $M$ ($0 \le M \le A$) denotes the 
number of excited atoms and $- A/2$ represents
the bottom energy level of the atomic system.  
Therefore, the collective atomic operators form a
$(A+1)$-dimensional representation of the
algebra su(2) corresponding to a spin $A/2$. The
case of a single resonant atom $(A=1)$ corresponds
to the well-known Jaynes-Cummings model.

For simplicity we shall restrict henceforth our 
attention to case of exact resonance between 
the atomic and the field frequency 
$\omega_{\mathrm{a}} = \omega_{\mathrm{f}}
\equiv \omega$. Since the field mode is described 
in the usual Fock space $| n \rangle_{\mathrm{f}}$, 
the natural bare basis for the total system is 
$|n, M \rangle \equiv  | n \rangle_{\mathrm{f}}
\otimes | M \rangle_{\mathrm{a}}$. However, 
it is straightforward to check that
\begin{equation}
[ H_0, H_{\mathrm{int}} ] =  0 ,
\end{equation}
so both are constants of motion. The Hamiltonian
$H_0$  (or, equivalently, the excitation number
$N$) determines  the total energy stored by 
the radiation field and the atomic system, which
is conserved by the interaction. This means that 
the appearance of $M$ excited atoms requires
the annihilation of $M$ photons. This allows us
to factor out $\exp(- i H_0 t)$ from the evolution
operator and drop it altogether. Hence, we
can relabel the total basis as
\begin{equation}
\label{basis}
|N-M, M \rangle \equiv  |N- M \rangle_{\mathrm{f}}
\otimes |M \rangle_{\mathrm{a}}  .
\end{equation}

In such a basis, the interaction Hamiltonian,
for a fixed value of $N$, is represented by
the tridiagonal matrix
\begin{equation}
H_{\mathrm{int}}^{(N)} = g
\left (
\begin{array}{ccccr}
0 & h_0 & 0 & . & \ldots \\
h_0 &  0 & h_1 & 0 & \ldots  \\
0 &  h_1 & 0 & h_2 & \ldots  \\
\vdots & \vdots & \vdots & \ddots &
\vdots 
\end{array}
\right ) ,
\end{equation}
with
\begin{equation}
h_M = 
\sqrt{(M+1) (N-M) (A -M)} .
\end{equation}
The dimension of this matrix depends
on whether $A > N$ or $A < N$, which 
are situations essentially different and must 
be  handled separately.

Let us assume that $A > N$ and initially all the  
atoms are unexcited. Then, $M = 0$ and the 
conservation of the number of excitations implies 
that only the states (\ref{basis}) with $0 \le M \le N$
take part in the dynamics. Thus, the dimension of 
the subspace is $N+1$.

On the contrary, when $A < N$  the number of 
initial photons is greater than the number of atoms 
and only the states (\ref{basis}) with $0 \le M \le A$ 
are  involved in the  evolution. The dimension is 
now  $A+1$

It is easy to check that, due to the properties
of the tridiagonal matrices, the eigenvalues
are distributed symmetrically with respect
to zero, with one eigenvalue equal to zero
if there are an odd number of them~\cite{Tanas91}.

To find the state evolution we shall need
the following matrix elements of the evolution 
operator 
\begin{equation}
C^N_{M^\prime M}(t) = 
\langle N-M^\prime, M^\prime |
\exp [- i H_{\mathrm{int}}^{(N)} t  ] 
|N-M, M \rangle, 
\end{equation}
which can be written as
\begin{equation}
C^N_{M^\prime M}(t) = \sum_{J=0}^{{\mathcal{D}}}
U_{M J} U^\dagger_{M^\prime J} 
\exp [- i \varepsilon_J^{(N)} t ]  ,
\end{equation}
where $U$ is the unitary matrix that diagonalizes 
the Hamiltonian and $ \varepsilon_I^{(N)}$ are the 
corresponding eigenvalues. In what follows we 
shall use the convention of denoting the dimension
of the Hamiltonian matrix $H_{\mathrm{int}}^{(N)}$ 
by ${\mathcal{D}}+1$, that is,
\begin{equation}
{\mathcal{D}}=  \min(N, A) .
\end{equation}

Now, let us assume that the initial field is
taken to be  in a coherent state $| \alpha 
\rangle_{\mathrm{f}}$ and that the
atomic state is initially prepared in an atomic
coherent state $| \zeta 
\rangle_{\mathrm{a}}$~\cite{Arecchi72,Perelomov86};
i.e.,
\begin{equation}
| \Psi (0) \rangle = 
| \alpha \rangle_{\mathrm{f}} \otimes 
| \zeta \rangle_{\mathrm{a}} ,
\end{equation}
where 
\begin{equation}
| \alpha \rangle_{\mathrm{f}} =
\sum_n Q_n \ | n \rangle_{\mathrm{f}} ,
\end{equation}
$Q_n$ being the Poissonian weighting
factor of the coherent state (with zero phase)
with mean number of photons $\bar{n}$
\begin{equation}
Q_n = \sqrt{e^{- \bar{n}} \frac{\bar{n}^n}{n!}} ;
\end{equation}
and 
\begin{eqnarray}
| \zeta \rangle_{\mathrm{a}} & = &
\frac{1}{\left ( 1 + |\zeta|^2 \right )^{A/2}} 
\sum_{M=0}^A \sqrt{\frac{A!}{M! (A-M)!}}
\zeta^M | M \rangle_{\mathrm{a}} \nonumber \\
& \equiv  & \sum_{M=0}^A A_M  \
| M \rangle_{\mathrm{a}}  ,
\end{eqnarray}
where the parameter $\zeta$ is normally
rewritten in terms of the spherical angles as
\begin{equation}
\zeta = - \tan(\vartheta/2)e^{- i \varphi} . 
\end{equation}
In other words, the initial state can be rewritten, 
taking into  account (\ref{basis}), as
\begin{equation}
\label{Init}
| \Psi (0) \rangle = \sum_{N,M} Q_{N-M} A_M
| N-M, M \rangle .
\end{equation}
With this initial condition the resulting state can
be recast as
\begin{eqnarray}
\label{evol}
| \Psi (t) \rangle & = & 
\exp(- i H_{\mathrm{int}} t)  | \Psi (0) \rangle \nonumber \\
& =  & \sum_{N=0}^\infty  
\sum_{M^\prime, M=0}^{{\mathcal{D}}} 
Q_{N-M} A_M \ C_{M^\prime M}^N (t) \ 
|N-M^\prime, M^\prime \rangle .
\end{eqnarray}
If the initial state is not of the same form, but it
has a decomposition with different amplitudes $A_M$
or $Q_n$, equation~(\ref{evol}) is still valid when the 
appropriate coefficients are taken.

\section{Relative-phase operator for the 
Dicke model}

In the spirit of our previous work on the relative phase 
for the Jaynes-Cummings model~\cite{phasJCM}, 
we shall describe the atom-field relative phase in 
terms of a polar decomposition of the complex
amplitudes. To this end, let us introduce 
the operators
\begin{eqnarray}
& X_+  =  a S_+ , \qquad 
X_-  =  a^\dagger S_- , &  \nonumber \\
& & \\
&  X_3  =  S_3 . & \nonumber 
\end{eqnarray}
These operators maintain the first commutation 
relation of su(2) in (\ref{ccrsu2}), $[X_3, X_\pm]  =  
\pm X_\pm$,  but the second one is modified in the 
following way:
\begin{equation}
[X_-, X_+] = P (X_3)  , 
\end{equation}
where $P (X_3)$ represents a second-order 
polynomial function of the operator $X_3$. This 
is a typical example of the so-called polynomial 
deformations of the algebra su(2). Without 
embarking us in mathematical subtleties, the 
essential point for our purposes here is that one 
can develop a theory in a very close analogy 
with the standard su(2) algebra. In particular, 
it is clear that the state $|N,0 \rangle $ plays the 
role of a \textit{vacuum state}, since
\begin{equation}
X_- | N, 0 \rangle = 0 .
\end{equation}
Then, we can construct invariant subspaces,
as in the usual theory of angular momentum, by
\begin{equation}
 | N-M, M \rangle = \frac{1}{\mathcal{N}} 
\ X_+^M |N, 0 \rangle ,
\end{equation}
where ${\mathcal{N}}$ is a normalization
constant. One can check that
\begin{equation}
X_+^{{\mathcal{D}}+1} | N, 0 \rangle = 0 ,
\end{equation}
confirming that the number of accessible states
is ${\mathcal{D}} +1$. 

In consequence, the whole space of the system 
can be split as the direct sum ${\mathcal{H}} = 
\oplus_{N = 0}^\infty {\mathcal{H}}_{N}$ 
of subspaces invariant under the action of the 
operators $(X_+, X_-, X_3)$ and each one of
them having a fixed number of excitations. These
independent subspaces do not overlap in the evolution, 
in such a way that if the initial state belongs to one
of them, it will remain in that subspace for all the
evolution.  

In each one of  these invariant  subspaces the 
operator $X_3$ is diagonal,  while $X_+$ and 
$X_-$ are ladder operators represented by 
finite-dimensional matrices. This suggests to 
introduce a polar decomposition in the form
\begin{eqnarray}
X_+ & = &  \sqrt{X_ + X_-} \ E \nonumber \\
& & \\
X_- & =  &   E^\dagger \sqrt{X_+ X_-} , \nonumber
\end{eqnarray}
where the \textit{radial} operator $\sqrt{X_+ X_-}$ 
is diagonal in the basis $ | N-M, M \rangle $: 
\begin{equation}
\sqrt{X_+ X_-} | N-M, M \rangle =
\sqrt{M(N-M+1)(A-M+1)} | N-M, M \rangle  ,
\end{equation}
and 
\begin{equation}
\label{lad}
[X_3, E] = E .
\end{equation}

We can guarantee now that the operator 
$E = \exp(i \Phi)$, representing the exponential 
of the relative phase, is unitary and commutes 
with the excitation number
\begin{eqnarray}
&  E E^\dagger = E^\dagger E = I , & \nonumber  \\
& & \\
& \ [E, N] = 0  . &  \nonumber
\end{eqnarray}
Thus, we may rather study its restriction to 
each invariant subspace ${\mathcal{H}}_N $, 
we shall denote by $E^{(N)}$. It is easy to check 
that the  action of the operator $E^{(N)}$ in each 
subspace is given by
\begin{eqnarray}
E^{(N)} | N-M, M \rangle & = & 
| N - (M+1), M+1 \rangle  , \nonumber  \\
& & \\
{E^{(N)}}^\dagger | N-M, M \rangle & = &
| N - (M-1), M- 1 \rangle . \nonumber  
\end{eqnarray}

Obviously, the action of $E^{(N)}$  and 
${E^{(N)}}^\dagger$ becomes undefined on 
the marginal states $| N -{\mathcal{D}}, {\mathcal{D}} 
\rangle$ and $| N, 0 \rangle$. Therefore, it is necessary
to add some conventions for closing the actions of 
these operators on the subspace ${\mathcal{H}}_N$. 
By analogy once again with the usual su(2) algebra, 
we shall  use standard cyclic conditions and impose
(up to global phase factors)
\begin{eqnarray}
E^{(N)} | N- {\mathcal{D}}, {\mathcal{D}} \rangle 
& = &   | N, 0 \rangle ,  \nonumber \\
& & \\
{E^{(N)}}^\dagger | N, 0 \rangle & = & 
| N- {\mathcal{D}}, {\mathcal{D}} \rangle. \nonumber
\end{eqnarray}
With these conditions, the operator $E^{(N)}$ 
can be expressed as
\begin{eqnarray}
E^{(N)} & = & \sum_{M=0}^{{\mathcal{D}}}
| N- (M+1), M +1 \rangle 
\langle N- M, M | \nonumber \\
& + & e^{ i ({\mathcal{D}}+1) \phi^{(N)} } 
| N, 0 \rangle \langle N-{\mathcal{D}}, {\mathcal{D}} | , 
\end{eqnarray}
$ \phi^{(N)}$ being an arbitrary phase. Note that the 
crucial extra term in this equation, which establishes 
the unitarity of $E^{(N)}$, is precisely based on the 
finite number of states. Therefore, in each invariant 
subspace ${\mathcal{H}}_N$ there are ${\mathcal{D}}+1$ 
orthonormal states  satisfying
\begin{equation}
E^{(N)} | \phi _{r}^{(N)} \rangle = e^{ i \phi _{r}^{(N)}}
| \phi_{r}^{(N)} \rangle ,
\end{equation}
with $r=0, \ldots, {\mathcal{D}}$. These states can 
be expressed as  
\begin{equation}
| \phi _{r}^{(N)} \rangle =
\frac{1}{\sqrt{{\mathcal{D}}+1}} 
\sum_{M=0}^{\mathcal{D}}
e^{i M \phi _{r}^{(N)}} | N-M, M \rangle ,
\end{equation}
and, by taking the same $2\pi $ window in each subspace,
we have
\begin{equation}
\phi_{r}^{(N)} = 
\phi _{0}+\frac{2\pi r}{{\mathcal{D}}+1} ,
\end{equation}
and $\phi_0$ is a fiducial or reference phase shift that 
can be arbitrarily chosen. The expression for $E$ on the 
whole space is
\begin{equation}
E = \sum_{N=0}^{\infty} E^{(N)}
= \sum_{N=0}^\infty 
\sum_{r=0}^{\mathcal{D}} 
| \phi _{r}^{(N)} \rangle  \
e^{ i \phi _r^{(N)}} \
\langle  \phi _r^{(N)} | ,
\end{equation}
and, since $E$ is unitary, it defines a Hermitian
relative-phase operator
\begin{equation}
\Phi = \sum_{N=0}^{\infty} \Phi^{(N)}
= \sum_{N=0}^\infty 
\sum_{r=0}^{\mathcal{D}} 
| \phi _{r}^{(N)} \rangle  \
\phi _r^{(N)} \
\langle  \phi _r^{(N)} | ,
\end{equation}
that, obviously, has discrete eigenvalues.
In the limit ${\mathcal{D}} \gg 1$, this 
spectrum becomes dense, as it might 
be expected. But, on the opposite limit,
one may be surprised to find that the
state $|0, 0 \rangle$ is a relative-phase
eigenstate (with arbitrary eigenvalue
$\phi_0$). While this may provide a
convincing argument that the theory
is unreasonable, we think that is not 
the case. The value of $\phi_0$ will 
not lead to any contradictions, because 
any  choice will lead to a consistent theory.
Our choice of this parameter says nothing
about Nature, it only  makes a statement
about our individual preference~\cite{Bjork}.

Note as well, that the relative-phase
eigenstates are maximally entangled
states. This has the consequence  that
the relative-phase operator has no classical
correspondence in the general case, not
even for highly excited states.

\section{Atom-field relative phase
in terms of absolute phases}

In the previous Section we have shown a clear
way of obtaining \textit{ab initio} the atom-field 
relative phase. In spite of this, one could still 
insist in describing  this variable in terms 
of the absolute phases of each subsystem. 
One must start then from previous descriptions 
of the field and atomic phases and manage them 
until getting the probability distribution for their difference.  
The goal of this section is to show that this way
of proceeding leads naturally to a positive 
operator-valued measure (POVM)~\cite{Helstrom76,Shapiro91}, 
and how such a POVM is precisely generated by 
the eigenstates of the relative-phase operator.

To this end, we shall adopt the elegant axiomatic
approach developed by Leonhardt
\textit{et al}~\cite{Leonhardt95} to describe 
the phase properties of both subsystems we 
are dealing with. To make the discussion as 
selfcontained as possible, we first briefly recall 
the essential ingredients of the general formalism.

When dealing with generic angle-action variables,  
one imposes that the complex exponential 
of the angle (denoted by $E$) and the action 
variable (denoted by $L_z$) satisfy 
[compare with Eq.~(\ref{lad})]
\begin{equation}
[L_z, E] = E .
\end{equation}
If $E$ were unitary, its action on the
basis of eigenstates of $L_z$ (denoted
by $|m \rangle$) will be as a ladder operator
\begin{equation}
E | m \rangle  = | m + 1 \rangle .
\end{equation}
The eigenstates of $E$ (denoted by 
$| \theta \rangle$) provide then an adequate 
description of the quantum angle~\cite{angle}. 
However, realistic measurements always involve 
extra  noise beyond that due to the intrinsic quantum 
fluctuations described by  canonical conjugation 
and it is essential to extend the canonical formalism 
by including fuzzy or unsharp generalizations of 
the ideal description provided by $E$~\cite{Hel74}.  
To this end we shall use POVMs, that are a set of linear
operators $\Delta (\theta)$ furnishing the correct
probabilities in any measurement process
through the fundamental postulate that  
\begin{equation}
P(\theta) = \mathrm{Tr} [ \rho \Delta (\theta) ] .
\end{equation}
The reality, positiveness, and normalization of 
$P(\theta)$ impose
\begin{equation}
\label{condpom}
\Delta^\dagger (\theta) = \Delta (\theta) ,
\quad
\Delta (\theta) \ge 0  ,
\quad
\int_{2 \pi} d\theta \ \Delta (\theta) = I,
\end{equation} 
but, in general, $\Delta (\theta)$ 
are not orthogonal projectors like in the standard 
measurements described by selfadjoint operators. 

In addition to these basic statistical conditions,
some other requirements must be imposed
to ensure that $\Delta (\theta)$ provides
a meaningful description of the angle as a canonically
conjugate variable with respect $L_z$ (even 
in the sense of a weak Weyl relation~\cite{Busch95}).
Then, we require
\begin{equation}
\label{req1}
e^{i \theta^\prime L_z}  \
\Delta (\theta) \ e^{- i \theta^\prime L_z}  
=  \Delta (\theta + \theta^\prime) ,
\end{equation} 
which reflects nothing but the basic feature
that an angle shifter is an angle-distribution
shifter.  This condition restricts the form
of the POVM to
\begin{equation}
\label{POM1}
\Delta (\theta) = \frac{1}{2 \pi} 
\sum_{n,m=0}^\infty b_{n,m} \
e^{i (m-n) \theta} \ | m \rangle \langle n| .
\end{equation} 
We must take also into account that a shift
in $L_z$ should not change the angle distribution. 
A shift in $L_z$ is expressed by the operator
$E$, since it shifts the distribution of $L_z$ by
one step. Therefore,  we require as well
\begin{equation}
\label{req2}
E   \ \Delta (\theta) \ E^\dagger   
=  \Delta (\theta ) ,
\end{equation} 
which, loosely speaking, is the physical translation 
of the fact that angle should be complementary to 
the action variable. This implies the invariance 
\begin{equation}
b_{n,m} = b_{n-m}  ,
\end{equation} 
that allows us to recast Eq.~(\ref{POM1}) as
\begin{equation}
\Delta (\theta) = \frac{1}{2 \pi} 
\sum_{\nu}^\infty b_{- \nu}
e^{- i \nu \theta} E^\nu ,
\end{equation} 
while conditions (\ref{condpom}) read now as
now
\begin{equation}
|b_\nu| \le 1, \quad
b_\nu^\ast = b_\nu .
\end{equation}

Expressing $E$ in terms of its eigenvectors
$| \theta \rangle$, we finally arrive at the 
general form of a POVM describing the angle 
variable and fulfilling the natural requirements~(\ref{req1}) 
and (\ref{req2}):
\begin{equation}
\label{genPOM}
\Delta (\theta) = \int_{2 \pi}
d \theta^\prime \ B(\theta^\prime) \
| \theta + \theta^\prime \rangle 
\langle \theta + \theta^\prime |,
\end{equation}
where
\begin{equation}
B(\theta) =  \frac{1}{2 \pi} 
\sum_{\nu =0}^\infty b_\nu e^{ i \nu \theta} .
\end{equation}
This convolution shows that this effectively
represents a noisy measurement, the function
$B(\theta)$ giving the resolution provided by this
POVM~\cite{angle}. 

Let us now focus on the phase properties 
of our two subsystems. Concerning the 
field phase,  the question has attracted the 
attention of physicists for almost as long as 
quantum mechanics has existed as a physical 
theory (for recent reviews see Ref.~\cite{revphas}). 
Nowadays, it seems indisputable that an operator 
representing the phase of a single-mode field in a
infinite-dimensional Hilbert space cannot exist~\cite{Fuji95} 
and the proper way to face up the problem involves
the use of a relative-state formalism. In spite 
of this serious drawback,  a variety of solutions 
have been proposed to circumvent the difficulties. 
Virtually all of them can be formulated within 
the POVM formalism discussed before, but  
with the role of $L_z$ being played by the 
number operator $a^\dagger a$. Then a 
number-shifter is expressed by the
Susskind-Glogower  phase operator~\cite{SG64}
(note that  we are not concerned about the problems
of $E$ as a phase operator here, we only use 
the number-shifter property) and the phase states are
\begin{equation}
| \theta_{\mathrm{f}} \rangle = \frac{1}{\sqrt{2 \pi}}
\sum_{n=0}^\infty e^{i n \theta_{\mathrm{f}}} 
| n \rangle_{\mathrm{f}} .
\end{equation}

On the other hand, the difficulties with hermiticity 
that phase operators encounter in the case of 
a single-mode field disappear for a two-level 
system.  In general, for the group SU(2) it is 
possible, by  working in the standard 
$(A+1)$-dimensional Hilbert space associated 
with the usual angular-momentum operators, 
to find a truly phase operator from a polar 
decomposition of the amplitudes 
$S_{\pm}$~\cite{Vourdas90,Ellinas90}. 
The procedure is quite similar to that followed in 
Sec.~III for the relative phase and, perhaps,  
the most striking consequence of this  
approach is that the atomic-dipole phase can take
only  $A +1$ different values, due to the
dimension of the atomic-state space.

Because of this particular behavior, one may 
think rather preferable to describe
the dipole phase by a POVM taking continuous
values in a $2 \pi$ interval, even though this
cannot lead to a standard operator description.
To this end, it suffices to note that the general
properties (\ref{req1}) and  (\ref{req2}) 
still hold, but the role of $L_z$ is played now
by $S_3$, and the ``Susskind-Glogower"
eigenstates of the shifter are now
\begin{equation}
\label{atSG}
| \theta_{\mathrm{a}}  \rangle =
\frac{1}{\sqrt{2 \pi}}
\sum_{M=0}^A e^{i M \theta_{\mathrm{a}} } 
| M \rangle_ {\mathrm{a}} .
\end{equation}

The joint probability distribution for atomic and 
field phases can be defined in a very natural way as
\begin{equation}
P(\theta_{\mathrm{a}}, \theta_{\mathrm{f}}) =
\mathrm{Tr} [ \rho
\Delta(\theta_{\mathrm{a}}, \theta_{\mathrm{f}}) ] ,
\end{equation}
with
\begin{equation}
\Delta(\theta_{\mathrm{a}}, \theta_{\mathrm{f}}) = 
\Delta^{\mathrm{a}} (\theta_{\mathrm{a}}) \otimes 
\Delta^{\mathrm{f}} (\theta_{\mathrm{f}}) .
\end{equation}
Our remaining task is to consistently
derive a POVM for the relative phase
$\phi= \theta_{\mathrm{a}} - \theta_{\mathrm{f}}$ 
from these expressions. This goal can be achieved
in many ways~\cite{PDPD}: for example, one can 
perform a  change of variables to express 
$\Delta(\theta_{\mathrm{a}}, \theta_{\mathrm{f}}) $ 
in terms of the phase sum and phase difference and 
then remove the phase-sum dependence by simple 
integration~\cite{angle}. Another possibility is to 
directly define the probability distribution for the 
relative phase as 
\begin{equation}
P( \phi) = \int_{2\pi} d\theta \ P(\theta, \theta + \phi) =
\mathrm{Tr} [ \rho \Delta (\phi) ] , 
\end{equation}
where
\begin{equation}
\Delta(\phi) =  \int_{2 \pi} d\theta \ 
\Delta (\theta, \theta + \phi) .
\end{equation}

For our purposes here, it is sufficient to note
that we must get the same values for any
periodic function of the relative phase whether 
we use the variable $\phi$ or  
$(\theta_{\mathrm{a}}, \theta_{\mathrm{f}})$. 
In consequence, we can impose that
\begin{equation}
\int_{2 \pi} d\phi \ P(\phi)  \ e^{i \nu \phi} =
\int_{2 \pi} d\theta_{\mathrm{a}} \ d\theta_{\mathrm{f}} \
P(\theta_{\mathrm{a}}, \theta_{\mathrm{f}}) \
e^{i \nu (\theta_{\mathrm{a}} - \theta_{\mathrm{f}})} .
\end{equation}

To proceed with we need to take a definite
choice for the corresponding POVMs. Concerning
the field phase, we recall that the Pegg-Barnett
formalism and many others embodying the concept
of phase as an observable canonically conjugate to 
photon number, lead to the POVM induced by 
the Susskind-Glogower phase states, 
namely~\cite{phasJCM}
\begin{equation}
\Delta^{\mathrm{f}}( \theta_{\mathrm{f}}) = 
|\theta_{\mathrm{f}} \rangle
\langle \theta_{\mathrm{f}} | .
\end{equation}
Motivated by this choice, we can use for the
atomic phase the finite-dimensional
translation of this POVM; i.e.,
\begin{equation}
\Delta^{\mathrm{a}} ( \theta_{\mathrm{a}}) = 
|\theta_{\mathrm{a}} \rangle
\langle \theta_{\mathrm{a}} | ,
\end{equation}
with $ |\theta_{\mathrm{a}} \rangle$ given
in Eq.~(\ref{atSG}). A simple calculation shows 
then that
\begin{equation}
\Delta (\phi) = \sum_{N=0}^\infty
|\phi^{(N)} \rangle
\langle \phi^{(N)} | , 
\end{equation}
where
\begin{equation}
\label{rpsg}
| \phi^{(N)} \rangle =
\frac{1}{2 \pi} 
\sum_{M=0}^{\mathcal{D}}
e^{i M \phi} | N-M, M \rangle .
\end{equation}
Therefore, we conclude that the POVM 
generated by the eigenstates of our relative phase 
operator is just the induced by other absolute-phase 
approaches (such as Susskind-Glogower or Pegg-Barnett),
when cast to the appropriate $2 \pi $ range. This is, 
in our view, another confirmation that the theory 
proposed works correctly.

\section{Relative-phase distribution function}

For any state, the information one can reap using
a measurement of some observable is given by
the statistical distribution of the measurement 
outcomes. For the relative phase, it seems natural
to define the probability distribution function of a
state described by the density matrix $\rho$ as
\begin{equation}
P(N, \phi_r, t) =  \langle  \phi _r^{(N)} | 
\rho (t) | \phi _{r}^{(N)} \rangle  .
\end{equation}
However, for \textit{physical states}~\cite{Barnett89}
(i.e., states for which finite moments of the
number operator are bounded) this expression 
will  converge to a simpler form involving
a continuous probability density we shall
write as
\begin{equation}
P(N, \phi, t) =  \langle  \phi ^{(N)} | 
\rho(t) | \phi ^{(N)} \rangle  ,
\end{equation}
where the vectors $| \phi^{(N)} \rangle $
defined in (\ref{rpsg}) lie in the subspace 
${\mathcal{H}}_N$ with total number of 
excitations $N$. In fact, this expression can
be interpreted as a joint probability distribution
for the relative phase and the total number
of excitations. From it, we can derive  the
distribution for the relative phase as
\begin{equation}
P(\phi, t) =  \sum_{N=0}^\infty P(N, \phi, t) ,
\end{equation}
while
\begin{equation}
P(N, t) =  \int_{2 \pi} d\phi \ P(N, \phi, t) 
\end{equation}
can be viewed as the probability distribution
of having $N$ excitations in the system.
These factorizations are an obvious consequence
of the fact that the relative phase and the excitation
number are compatible.

For the general initial state of Eq.~(\ref{Init})
and the evolution given by Eq.~(\ref{evol})
we have
\begin{equation}
P(N, \phi, t) =  | \langle  \phi ^{(N)} | \Psi(t) \rangle |^2 ,
\end{equation}
which, through direct calculation, gives
\begin{equation}
P(N, \phi, t) =  \frac{1}{2 \pi} \left |
\sum_{M^\prime, M=0}^{\mathcal{D}}
Q_{N-M} A_{M}\ 
C_{M^\prime  M}^{N}(t) e^{i M^\prime \phi} \right |^2 ,
\end{equation}
and then we arrive at the total relative-phase
probability distribution:
\begin{equation}
\label{Pdrp}
P(\phi, t) =  \frac{1}{2 \pi} \sum_{N=0}^\infty
\left | \sum_{M^\prime, M=0}^{\mathcal{D}}
Q_{N-M} A_M \ C_{M^\prime  M}^{N}(t)  
e^{i M^\prime \phi} \right |^2 .
\end{equation}
This is our basic and compact result which we use
to analyze the evolution of the phase properties of
the Dicke model.

In Fig.~1 we have numerically evaluated this
distribution $P(\phi, t)$ as a function of $\phi$ and
the rescaled adimensional time $gt$, for the case
when all the atoms are initially unexcited and the 
field is in a coherent state with various values of 
the mean number of photons $\bar{n}$. In all the 
cases, when  $\tau=0$ we have that 
$C_{M^\prime M}^N(0) = \delta_{M^\prime M}$ 
and therefore
\begin{equation}
P(\phi, t=0) =  \frac{1}{2 \pi} \sum_{N=0}^\infty
\left | \sum_{M=0}^{\mathcal{D}}
Q_{N-M} A_M \ e^{i M \phi} \right |^2 .
\end{equation}
In particular, when all the atoms are initially
unexcited only the coefficient $A_0$  survives 
and the previous expression  reduces to 
\begin{equation}
P(\phi, t=0) =  \frac{1}{2 \pi} .
\end{equation}
This flat distribution reflects the fact that the 
random phase of the dipole in such states induces
a uniform distribution centred at $\phi_0$. 
At this respect, it is interesting to notice that 
classically the Lorentz model at resonance
predicts  for the relative phase values of $\pm \pi/2$. 
It turns out that this is also a possible choice  
to fix the reference phase 
$\phi_0$ in the quantum  description. For 
simplicity, in all the graphics we have 
chosen $\phi_0$ as the origin $0$.

Two quite different behaviors are evident
from these graphics. The first occurs
in the weak-field region~\cite{MacKoz,Heidman85,Butler86}, 
when the number of excitations in the system is much smaller
than the number of the atoms, $N \ll A$.
If, for simplicity, we assume that all the
atoms are unexcited and the average number
of photons in the initially coherent field is small,
say $\bar{n}  \sim 1$, then we can retain 
only the dominant terms in Eq.~(\ref{Pdrp}), getting
\begin{equation}
P(\phi, t) \simeq  \frac{1}{2 \pi} 
\{ 1 + \bar{n} 
[ |C_{00}^1 |^2 +
| C_{01}^1 |^2 
+ 2 \mathrm{Re}  (
C_{00}^1{C_{01}^1}^\ast e^{i \phi}  )
 ] \} e^{- \bar{n}}  .
\end{equation}
We see that, due to the periodic temporal dependence
of the terms $C_{M^\prime M}^N(t)$, this
distribution is oscillatory for all times, which is  
corroborated numerically in Fig.~1.a.

The second  (and perhaps more interesting) case 
corresponds to the strong-field 
region~\cite{Chumakov,Drobny93,Knight93}, 
when the initial number of photons is much larger than
the number of atoms $A \ll N$.  Then,
following the ideas of Ref.~\cite{Chumakov} one 
can show that the coefficients  $C_{M^\prime M}^N (t)$ 
can be approximated, up to order $A/\sqrt{\bar{n}}$, by 
\begin{equation}
C_{M^\prime M}^N (t) \simeq
d_{M^\prime M}^A (- \Omega_N t) ,
\end{equation}
where 
\begin{equation}
\Omega_N = 2 g \sqrt{N - A/2 + 1/2} ,
\end{equation}
and $d_{M^\prime M}^A$ are Wigner $d$ 
functions~\cite{Vilenkin91}, which are defined as 
the matrix elements for finite  rotations by 
operators from SU(2) group representations
\begin{equation}
d_{M^\prime M}^A (\vartheta) =
d_{M M^\prime}^A (\vartheta) =
\langle M^\prime | e^{i \vartheta S_x} | M \rangle ,
\end{equation}
where $M, M^\prime=0, 1, \ldots, A$. The point
now is that essentially only one subspace of dimension
$A + 1$ dominates the dynamics. Moreover, a simple
calculation using the explicit form of these $d$
functions, gives
\begin{equation}
P(\phi, t) =  \frac{1}{2 \pi} \sum_{N=0}^\infty
Q_N^2 \left | \sum_{M=0}^A
\sqrt{\frac{A!}{(A-M)! M!}} 
\tan^M ( \Omega_N t/2) e^{i M (\phi - \pi/2)}
\right |^2 \cos^{2A}(\Omega_N t/2) ,
\end{equation}
where we have assumed that all the atoms 
are initially unexcited. When $A \gg 1$ and when
oscillations are well resolved, one can perform an
expansion of the square root getting
\begin{equation}
P(\phi, t) =\sqrt{\frac{A}{2 \pi}} \sum_N \Phi_N (t) \ 
\exp \left [ - \frac{A}{2} (\phi - \pi/2 + \delta_N )^2
\right ] ,
\end{equation}
where $\Phi_N (t)$ is a function of time of complicated
structure that accounts for the collapses and
revivals and that is of little interest for our purposes
here, and $\delta_N = \arg[\tan(\Omega_N t/2)]$.
Now, it is clear that, since $\delta_N $ takes only
the values $0$ and $\pi$, the previous Gaussian
distributions tend to have two peaks at
$\phi = \pm \pi/2$, in agreement with the
classical expectations.  The presence of collapses 
and revivals are evident in Fig.~1.b,  which 
confirms previous numerical and analytical evidence.
The well-known nearly time-independent behavior
in the time windows between collapse and revival 
is also clear. As we can see, the distribution tends to be
randomized in the evolution, although keeping 
these two peaks at $\pm \pi/2$.

In the intermediate region, when $N \sim A$,
the behavior is more complex, as shown in 
Fig.~1.c, and no analytical approximations are
available.

For the particular case of the Jaynes-Cummings model
one can diagonalize exactly the Hamiltonian in
each subspace ${\mathcal{H}}_N$, obtaining
the well-known dressed states~\cite{Cohen89}, that turn
to be trapping states~\cite{Gea91}; i.e., the atomic population 
$\langle S_3 (t)  \rangle$  remains constant in 
spite of the existence of both the radiation field 
and atomic transitions~\cite{Cirac}. These states play 
a  fundamental role in that model, so it seems
interesting to analyze the corresponding problem
for the case of the Dicke model. In the strong-field limit  
one can make the replacement $a \rightarrow \alpha =
\sqrt{\bar{n}} \ e^{i \theta}$ and the interaction Dicke 
Hamiltonian becomes proportional to the operator
\begin{equation}
H_{\mathrm{cl}} =  
\left ( e^{i \theta} S_+  + e^{- i \theta} S_ - \right )  , 
\end{equation}
where the phase of the classical field has been chosen 
to coincide with the phase of the initial coherent state
of the field.  The semiclassical atomic states 
are defined now as eigenstates of  
$H_{\mathrm{cl}}$ taking this phase as zero:
\begin{equation}
2 S_x |\underline{P} \rangle_{\mathrm{a}} = 
\Lambda_P |\underline{P} \rangle_{\mathrm{a}} , 
\end{equation}
with $\Lambda_P = A - 2 P$ and $P=0, 1, \ldots, A$.

Following Ref.~\cite{Chumakov}, we shall call 
\textit{factorized states} those states for which  the initial 
field is taken to be in  a  strong coherent  state 
$| \alpha \rangle_{\mathrm{f}}$  and the  atomic 
system is initially prepared in a semiclassical 
atomic state $|\underline{P} \rangle_{\mathrm{a}}$.
For such states, the total wavefunction of the
system can be approximately written as a product
of its field and atomic parts
\begin{equation}
\label{semtra}
| \Psi(t) \rangle \simeq 
|\underline{P}(t) \rangle_{\mathrm{a}} \otimes
| \alpha (t) \rangle_{\mathrm{f}}
\end{equation}
with
\begin{eqnarray}
|\underline{P}(t) \rangle_{\mathrm{a}} & = &
\exp \left [ - i \frac{ \Lambda_P (S_3 + A/2)}
{2 \sqrt{\bar{n} - A/2 + 1/2}} g t  \right ] 
|\underline{P} \rangle_{\mathrm{a}} \nonumber \\
& & \\
| \alpha (t) \rangle_{\mathrm{f}} & = & 
\exp \left [ - i  \Lambda_P 
\sqrt{a^\dagger a - A/2 + 1/2} \ g t   \right ] 
| \alpha  \rangle_{\mathrm{f}} ,  \nonumber 
\end{eqnarray}
and one can verify that they are also
(approximately) trapping states. For these 
states, one can find after a  simple calculation, 
\begin{equation}
P(\phi, t ) =  \frac{1}{A+1} 
\left | \sum_M {}_{\mathrm{a}}\langle M |
\underline{P} \rangle_{\mathrm{a}} e^{i M \phi} 
\right |^2 .
\end{equation}
The probability distribution is time independent due
to the factorization (\ref{semtra}). From the 
arguments in Ref.~\cite{Chumakov}, one infers
that this factorization holds up to times 
$g t \sim \sqrt{\bar{n}}$ (which can be very 
long times, in this limit) and with an accuracy 
in the coefficients of the order of $A/\sqrt{\bar{n}}$.

Moreover, using the  properties of the semiclassical 
atomic states and assuming $A \gg 1$, one can replace
the sum by an integral, obtaining finally
\begin{equation}
P(\phi, t) \simeq  \sqrt{\frac{A}{2 \pi}} \
e^{- A \phi^2/2} ;
\end{equation}
i.e., a Gaussian  independent of time. In Fig.~2
we have plotted the probability distribution obtained
from a numerical computation of Eq.~(\ref{Pdrp}), 
showing this quite remarkable behavior, except for 
the presence of very small (almost inappreciable) 
oscillations superimposed.

To gain more physical insight in  these behaviors,
in Fig.~3 we have plotted the evolution of the 
mean value of $\langle \sin \Phi \rangle$ for
various values of $N$, confirming the previous
physical discussion.

To conclude, let us consider the Dicke model
in the large-detuning limit; which is usually
known as the dispersive limit. More specifically,
we are in the case when 
\begin{equation}
\Delta \gg g \sqrt{\bar{n}+1} A.
\end{equation}
Then, following the procedure developed in
Ref.~\cite{small}, the interaction Hamiltonian
in Eq.~(\ref{Hamex}) cn be replaced by
the effective  Hamiltonian 
\begin{equation}
\label{Hamef}
H_{\textrm{eff}} = \Delta \ S_3 +
\lambda (2 a^\dagger a + 1) S_3 +
\lambda (C - S_3^2) ,
\end{equation}
where 
\begin{equation}
C = \frac{A}{2} \left ( \frac{A}{2} + 1
\right ) , \qquad \lambda = \frac{g^2}{\Delta} .
\end{equation}
The obvious advantage of this Hamiltonian 
is that it is diagonal and allows for a compact 
analytical expression for the coefficients
$C^N_{M^\prime M}(t) $ as 
\begin{eqnarray}
C^N_{M^\prime M}(t) & = & \delta_{M^\prime M}
\exp \left ( - i  t \left \{ \Delta (M- A/2) + 
\lambda \left [ 2 (N- M)  + 1 \right ] 
(M- A/2)  \right . \right .   \nonumber \\
& +& \left . \left .  \lambda  \left [ C -
(M- A/2)^2 \right ] \right \} \right ) .
\end{eqnarray}
When the atoms are initially unexcited  or excited
(or, more generally, when $A_M = \delta_{MK}$)
and for any initial state of field, we have
\begin{equation}
P(\phi, t) = \frac{1}{2 \pi},
\end{equation}
for all the times. 

For an arbitrary initial state of the atomic system and 
the field we get 
\begin{equation}
\label{shcat}
P(\phi, t) = \frac{1}{2\pi }  \sum_{N=0}^\infty 
\left | \sum_{M=0}^{A} Q_{N-M} A_M
e^{-i f_M^Nt} e^{i M \phi }\right |^2 ,
\end{equation}
with 
\begin{equation}
\label{fMN}
f_M^N = 2NM \lambda + 
[ \Delta +\lambda (2A+1) ] M- 
3 \lambda M^2.
\end{equation}
Since (\ref{Hamef}) is quadratic in the population 
inversion operator $S_3$ and is, therefore, analogous 
to the Hamiltonian quadratic in the number operator 
of a single-mode field propagating through a Kerr 
medium, one could expect~\cite{Agarwal97} that 
the evolution of coherent atomic states in the 
dispersive limit of the Dicke model  leads to the 
generation of Schr\"{o}dinger cat states. This 
superposition reaches the most pure form for 
initial number field states (in particular, the 
vacuum state minimizes the atomic 
entropy~\cite{saavedra98}). 

The situation with the relative-phase distribution is quite 
different. It is easy to see, for example, that if the field 
is prepared initially in a number state $|k\rangle $  then 
$Q_n= \delta_{kn}$  and the relative phase distribution is 
flat. Nevertheless, for initial atomic and field coherent 
states the relative-phase distribution splits for some 
special times into several humps. These catlike states, 
according to (\ref{fMN}), appear at 
times $\tau = \lambda t = \pi/6 \ (\bmod 2\pi)$.
To confirm analytically this behavior,
we expand Eq.~(\ref{shcat}) when initially
we have strong coherent states for both field
and atoms, with  $\bar{n} \gg A \gg 1$. By replacing
once again the sum by integrals, one easily gets
\begin{equation}
P(\phi, t= \pi/6\lambda ) = \sqrt{\frac{A}{8\pi }} 
\left \{ e^{- \left [ \phi - \pi  \phi_{\bar{n}} /(3 \lambda) \right ]^2 A/2} 
+ e^{- \left [ \phi + \pi - \pi \phi_{\bar{n}}/(3 \lambda) \right ]^2 A/2}
\right \} ,
\end{equation}
where
\begin{equation}
\phi_{\bar{n}} =  
2 \bar{n} \lambda +  A +
\lambda (2A+1) ,
\end{equation}
and all the phases must be understood $\bmod(2\pi)$. The
two separated Gaussians indicates the presence of
two humps and, therefore, the presence of catlike states.
To further confirm this, in Fig.~4 we have computed 
numerically the distribution function $P(\phi, t= \pi/6\lambda )$ 
at the times predicted by the theory. The graphic clearly
demonstrates the presence of the two-component state,
according to our previous considerations.

\section{Conclusions}

In this paper we have investigated an 
appropriate operator for the quantum
description of the relative phase in the 
Dicke model. We have used a proper 
polar decomposition of the corresponding 
field amplitudes, much in the spirit of our 
previous work on the subject. This polar 
decomposition has been justified on physical 
grounds, as well as using the theory of polynomial
deformations of su(2).

The eigenvalue spectrum of this operator
is discrete, as it happens for the polar
decomposition corresponding to two 
field modes. From these eigenstates we
have obtained the probability distribution
for the relative phase and we have
studied its time evolution. For the weak-field
region, the behavior is essentially oscillatory,
while for the strong-field region, the relative
phase tends to be randomized in the evolution,
although showing collapses and revivals.
For both limiting regions, we have developed
analytical approximations that have allowed
us an easy physical interpretation of some 
remarkable phenomena.

\begin{figure}
\caption{Gray-level  contour plot of the
probability distribution $P(\phi, t)$
as a function of $\phi$ and the rescaled
time  $gt$ for the case of $A=5$ atoms 
initially unexcited and the field in a 
coherent state with the following values
of the mean number of photons:
a) $\bar{n}=1$ (weak field), 
b) $\bar{n}=50$ (strong field), and 
$\bar{n}=5$ (intermediate field).}
\end{figure}

\begin{figure}
\caption{Probability distribution 
function $P(\phi, t)$ as a function of 
$\phi$ and the rescaled time $gt$ 
for the case of a factorized state with 
$A=3$.  The atomic coherent state
has $\vartheta = \pi/2$ and 
$\varphi =0$ and the field
state has $\bar{n}=20$.}
\end{figure}

\begin{figure}
\caption{Plots of $\langle \sin \phi \rangle$
versus $gt$ for the same values of 
$\bar{n}$ as in Fig.~1 (from top to bottom).}
\end{figure}

\begin{figure}
\caption{Probability distribution 
function $P(\phi, t)$ as a function of 
$\phi$ for the time $\lambda t = \pi/6$ 
for the case of an atomic coherent state
$(A=5)$ with $\vartheta = \pi/2$ and  $\varphi =0$ 
and a field coherent state with $\bar{n}=10$. The
presence of the two humps corroborate the
presence of a catlike state.}
\end{figure}

\begin{references}

\bibitem{Allen87}
L. Allen and J. H. Eberly,
\textit{Optical Resonance and Two-Level Atoms}
(Dover, New York, 1987).

\bibitem{Dicke54}
R. Dicke,
Phys. Rev. \textbf{93}, 99 (1954).

\bibitem{Tavis68}
M. Tavis and F. W. Cummings,
Phys. Rev. \textbf{170}, 379 (1968).

\bibitem{Stroud70}
C. P. Stroud and E. T. Jaynes,
Phys. Rev. A \textbf{1}, 106 (1970);
\textbf{2}, 1613 (1970).

\bibitem{Kumar70}
S.  Kumar and C. L. Mehta,
Phys. Rev. A \textbf{2}, 1573 (1970).

\bibitem{Cives99}
A. Cives-Esclop, A. Luis, and L. L. S\'anchez-Soto,
J. Mod. Opt. \textbf{46}, 639 (1999).

\bibitem{Meystre91}
P. Meystre and M. Sargent III,
\textit{Elements of Quantum Optics}
(Springer-Verlag, Berlin, 1991).

\bibitem{Ashcroft96}
N. W. Ashcroft and N. D. Mermin,
\textit{Solid State Physics}
(Saunders College, Philadelphia, 1996).

\bibitem{shg}
J. Delgado, A. Luis, L. L. S\'anchez-Soto, and A. B. Klimov, 
J. Opt. B: Quantum Semiclass. Opt. \textbf{2}, 33 (2000).

\bibitem{relpha}
A. Luis and  L. L. S\'anchez-Soto,
Phys. Rev. A  \textbf{48}, 4702 (1993);
L. L. S\'anchez-Soto and A. Luis, 
Opt. Commun. \textbf{105}, 84 (1994);
A. Luis, L. L. S\'anchez-Soto, and R. Tana\'s,
Phys. Rev. A {\bf 51}, 1634 (1995).

\bibitem{deformed}
V. P. Karassiov, 
J. Phys. A \textbf{27}, 153 (1994);
V. P. Karassiov and A. B. Klimov, 
Phys. Lett. A \textbf{191}, 117 (1994);
C. Quesne,
J. Phys. A \textbf{28}, 2847 (1995);
B. Abdessalam, J. Beckers, A. Chakrabart,
and N. Debergh,
\textit{ibid.} \textbf{29}, 3075 (1995);
N. Debergh,
\textit{ibid.} \textbf{31}, 4013 (1998).

\bibitem{Tanas91}
R. Tana\'s, Ts. Gantsog, and R. Zawodny,
Quantum Opt. \textbf{3}, 221 (1991).

\bibitem{Arecchi72}
F. T. Arecchi,  E. Courtens, R. Gilmore, and H. Thomas,
Phys. Rev. A \textbf{6}, 2211 (1972).

\bibitem{Perelomov86}
A. Perelomov,
\textit{Generalized Coherent  States and
their Applications}
(Springer-Verlag, Berlin, 1986).

\bibitem{phasJCM}
A. Luis and  L. L. S\'anchez-Soto,
Phys. Rev. A  \textbf{56}, 994 (1997).

\bibitem{Bjork}
G. Bj\"{o}rk, A. Trifonov, T. Tsegaye, and 
J. S\"{o}derholm,
Quantum Semiclass. Opt. \textbf{10}, 705 (1998);
A. Trifonov, T. Tsegaye, G. Bj\"{o}rk, 
J. S\"{o}derholm, E. Goobar, M. Atat\"{u}re,
and A. V. Sergienko,
J. Opt. B: Quantum Semiclass. Opt. 
\textbf{2}, 105 (2000).

\bibitem{Helstrom76}
C. W. Helstrom,
\textit{Quantum Detection and Estimation Theory}
(Academic, New York, 1976).

\bibitem{Shapiro91}
J. H. Shapiro and S. R. Shepard,
Phys. Rev. A  \textbf{43}, 3795 (1991).

\bibitem{Leonhardt95}
U. Leonhardt, J. A. Vaccaro, 
B. B\"{o}hmer, and H. Paul,
Phys. Rev. A  \textbf{51}, 84 (1995).

\bibitem{angle}
A. Luis and  L. L. S\'anchez-Soto,
Eur. Phys. J. D  \textbf{3}, 195 (1998).


\bibitem{Hel74}
C. W. Helstrom, 
Int. J. Theor. Phys. \textbf{11}, 357 (1974);
J. H. Shapiro, S. R. Shepard, and N. C. Wong,
Phys. Rev. Lett.  \textbf{62}, 2377 (1989);
S. L. Braunstein, C. M.  Caves, and  G. J. Milburn,
Phys. Rev. A \textbf{43}, 1153 (1991);
S. Stenholm,
Ann. Phys. (NY) \textbf{218}, 233 (1992);
U. Leonhardt  and H. Paul,   
J. Mod. Opt. \textbf{40}, 1745 (1993). 

\bibitem{Busch95}
P. Busch, M. Grabowski, and P. Lahti,
Ann. Phys. (NY) \textbf{237}, 1 (1995).


\bibitem{revphas}
J. Bergou and B. G. Englert,
Ann. Phys. (NY) \textbf{209}, 470 (1991);
Phys. Scr.  \textbf{T48}, 1 (1993), special
issue on quantum phase;
A. Luk\v{s} and V. Pe\v{r}inova,
Quantum Opt. \textbf{6}, 125 (1994);
R. Lynch,
Phys. Rep. \textbf{256}, 367 (1995);
R. Tana\'s, A. Miranowicz, and Ts. Gantsog,
Prog. Opt. \textbf{36}, 161 (1996);
A. Luis and L. L. S\'anchez-Soto,
Prog. Opt. {\bf 41}, 421 (2000).

\bibitem{Fuji95}
K. Fujikawa,
Phys. Rev. A \textbf{52}, 3299 (1995).

\bibitem{SG64}
L. Susskind and J. Glogower, 
Physics (NY) \textbf{1}, 49 (1964).

\bibitem{Vourdas90}
A. Vourdas,
Phys. Rev. A \textbf{41}, 1653 (1990).

\bibitem{Ellinas90}
D. Ellinas, 
J. Math. Phys. \textbf{32}, 135 (1990).

\bibitem{PDPD}
A. Luis and L. L. S\'anchez-Soto,
Phys. Rev. A {\bf 53}, 495 (1996)

\bibitem{Barnett89}
S. M. Barnett and D. T. Pegg,
J. Mod. Opt. {\bf 36}, 7 (1989)

\bibitem{MacKoz}
M. Kozierowski, A. A. Mamedov,
and S. M. Chumakov,
Phys. Rev. A {\bf 42}, 1762 (1990);
M. Kozierowski, S. M. Chumakov,
J. Swiatlowski, and A. A. Mamedov,
\textit{ibid.} {\bf 46}, 7220 (1992).

\bibitem{Heidman85}
A. Heidmann, J. M. Raimond,
and S. Reynaud, 
Phys. Rev. Lett. {\bf 54}, 326 (1985).

\bibitem{Butler86}
M. Butler and P. D. Drummond,
Opt. Acta {\bf 33}, 1 (1986).

\bibitem{Chumakov}
S. M. Chumakov, A. B. Klimov, and
J. J. S\'anchez-Mondrag\'on,
Phys. Rev. A {\bf 49}, 4972 (1994);
Opt. Commun. {\bf 118}, 529 (1995);
A. B. Klimov and S. M. Chumakov,
Phys. Lett. A {\bf 202}, 145 (1995);
J. C. Retamal, C. Saavedra,
A. B. Klimov, and S. M. Chumakov,
Phys. Rev. A {\bf 55}, 2413 (1997).

\bibitem{Drobny93}
G. Drobn\'y and I. Jex,
Opt. Commun. {\bf 102}, 141 (1993).

\bibitem{Knight93}
P. L. Knight and B. W. Shore,
Phys. Rev. A {\bf 48}, 642 (1993).

\bibitem{Vilenkin91}
N. Vilenkin and A. Klimyk,
\textit{Representation of Lie Groups
and Special Functions}
(Kluwer Academic, Dordrecht, 1991),
Vols. 1-3.

\bibitem{Cohen89}
C. Cohen-Tannoudji, J. Dupont-Roc,
and G. Grynberg, 
\textit{Photons and Atoms: Introduction
to Quantum Electrodynamics}
(Wiley, New York, 1989).

\bibitem{Gea91}
J. Gea-Banacloche,
Phys. Rev. A \textbf{44}, 5913 (1991).

\bibitem{Cirac}
J. I. Cirac and L. L. S\'anchez-Soto,
Phys. Rev. A \textbf{42}, 2851 (1990);
Opt. Commun. \textbf{80}, 67 (1990).

\bibitem{small}
A. B. Klimov and L. L. S\'anchez-Soto,
Phys. Rev. A \textbf{61}, 063802 (2000).

\bibitem{Agarwal97}  
G. S. Agarwal, R. R.Puri,  R. P. Singh, 
Phys. Rev. A \textbf{56}, 2249 (1997).

\bibitem{saavedra98}  
A. B. Klimov and C. Saavedra, 
Phys.Lett. A, \textbf{247}, 14 (1998).


\end{references}
\end{document}